\documentclass{elsart}

\usepackage{natbib}

\usepackage{graphics}
\usepackage{graphicx}
\usepackage{epsfig}

\usepackage{amssymb}

\begin{document}

\begin{frontmatter}




\title{The evolution of Radio Loud Quasar host galaxies: AO observations at z $\sim$ 3.}


\author[padova]{R. Falomo}
\author[turku]{J.K.  Kotilainen} 
\author[eso]{R. Scarpa} 
\author[como]{A. Treves}
\author[milano]{M. Uslenghi}

\address[padova]{INAF-Osservatorio di Padova, Italy} 
\address[turku]{Tuorla Observatory, Turku, Finland}
\address[eso]{ESO, Santiago, Chile}
\address[como]{University of Insubria, Como, Italy}
\address[milano]{IASF-INAF- Milano}

%

\begin{abstract}

We report on ESO--VLT  adaptive optics imaging of one radio-loud quasar at z $\sim$ 3.
In spite of the large distance of the object we are able to detect its 
surrounding extended nebulosity 
the properties of which are consistent with an underlying massive galaxy of 
M$_K$ $\sim$ --27 and effective radius R$_e$ = 7 kpc.
As far as we know this is the clearest detection of a radio loud quasar host at high redshift. 
The host luminosity is indicative of the existence of massive spheroids even at this
early cosmic epoch. The host luminosity is 
about 1 magnitude fainter than the expected value based on the average trend 
of the host galaxies of RLQ at lower redshift. 
The result, which however is based on a single object, 
suggests that at z $\sim$ 3  there is a deviation from a luminosity--redshift dependence 
regulated only by passive evolution. 
\end{abstract}


\end{frontmatter}


\section{Introduction}
\label{intro}

It is well known that at low redshift quasars are hosted in otherwise
normal luminous (and massive) galaxies \citep{bahcall97,dunlop03,pagani03} characterized by a 
conspicuous spheroidal component that becomes dominant in radio loud
objects (RLQ). 
These galaxies appear to follow the same relationship between the bulge luminosity 
and the mass of the central black hole observed in nearby inactive elliptical galaxies \citep{ferrarese02}. 
If this link keeps also at higher redshift the observed population of high z quasars traces  the existence 
of $ \sim 10^9$ M$_\odot$  super massive BHs and massive spheroids at very early ($<$ 1 Gyr) cosmic epochs 
\citep{fan01,fan03,willott03}.   

Indeed this idea seems also supported by the discovery of molecular gas and metals 
in high z quasars \citep{bertoldi03,freudling03}, that are suggestive of galaxies with strong star formation.
In this context it is therefore important to push as far as possible in redshift the direct
detection and characterization of QSO host galaxies. 
In particular, a key point is to probe the QSO host properties at
epochs close to (and possibly beyond) the peak of quasar activity (z $\sim$ 2.5).  

Till few years ago, due to the severe observational difficulty, the properties of quasar hosts at high redshift
were very poorly known (e.g. see the pioneering papers by 
\citet{hutchings95,lehnert92,lowenthal}. Uncertain
or ambiguous results, were produced because based on inadequate quality of the images (modest resolution; 
low S/N data; non optimal analysis).

In this work we present first results of a program aimed at imaging the host galaxies of quasar at z $>$ 2 using
adaptive optics at 8m class telescopes. Throughout this work we use H$_0$ = 70 km s$^{-1}$ Mpc$^{-1}$, $\Omega_m$
= 0.3, and $\Omega_\Lambda$ = 0.7.

\section{Adaptive optics imaging of distant QSO}

In order to characterize the properties of high z quasar hosts it is of
fundamental importance to combine high spatial resolution (narrow PSF) with very good sensitivity to detect and
measure the faint nebulosity surrounding the bright QSO nuclei. 
Adaptive optics (AO) opened a new window in this field, though the first generation of AO systems at 4m class 
telescopes enabled the detection of details on the host at low $z$ they 
did not allow much improvements for distant quasars 
\citep{hutchings98,hutch99,marquez01,lacy02}.

It was the recent introduction of sophisticated AO systems at 8m class telescopes that, for the first time,  
provided the spatial resolution and the adequate sensitivity for pushing the detection of QSO hosts at
z$>$2, and it did not take long for new results to appear in the
literature. \citet{croom04} using the AO system at the Gemini North telescope 
were able to resolve and characterize one
radio quiet quasar (RQQ) at z=1.93, finding for the host an absolute
magnitude of M$_K$ = -27.3. In a pilot program at the ESO VLT equipped with the AO system (NACO), 
\citet{falomo05} resolved a radio loud quasar at z $\sim$ 2.5. The absolute
magnitude of the host was found M$_K$ = --27.6. In both
cases the host galaxy luminosity appears to be consistent with the
trend followed at lower redshift \citep{falomo04,kotilainen05}.
RQQ and RLQ  host luminosity--redshift dependence follows that of massive spheroids 
undergoing simple passive evolution. Up to z $\sim$ 2 the average host luminosity is 
about 4L$^*$ for RLQ and 0.7 mag fainter for RQQ. 

In order to investigate  the properties of quasar hosts at z $>$ 2.5 and to explore 
the region near the peak of QSO activity we have carried out a new program to secure K band images 
of quasars in the redshift range 2 $<$ z $<$ 3 using the AO system at ESO VLT.  

\section{Object selection, observations and data analysis}

Adaptive optics systems employing natural guide stars imply that only
the targets that are sufficiently  close to relatively bright stars (used as
reference for AO correction) can be actually observed.  
In order to find an adequate number of targets for AO observations 
we searched the Veron \& Cetty-Veron (2003) catalog of quasars 
to collect suitable objects in the redshift range 2$<$ z $<$ 3 and
$\delta <$ 0, for sources having a star brighter than V=14 within 30 arcsec. 
Under these conditions the system is expected to deliver
images of Strehl ratio better than $\sim$ 0.2 when the seeing
is $<$0.6 arcsec. Among these candidates we also searched for objects that have 
other stars in the field usable to derive the PSF. 
This produced a list of twenty objects  only three of which are radio loud.
Here we present the results for one RLQ at z $\sim$ 3 : WGA J0633.1-2333 (
B = 21.5 ; z = 2.928, Perlman et al. 1998 ).


Two K-band images  were obtained using NAOS -- CONICA \citep{rousset02,lenzen98}, the
 AO system on the VLT at the European Southern Observatory (ESO) in Paranal (Chile). 
 The detector used ( Aladdin InSb 1024x1024 pixels) provides a field of view of 56x56 
 arcsec with a sampling of 54 mas/pixel. The object was observed using a jitter 
 procedure and individual exposures of 2 minutes per frame, for a total integration time of 48
minutes (see Figure 1). 





\section{Results }      



We performed a detailed modeling of the image of the object 
using a new software specifically designed to perform two dimensional model fitting of QSO images 
AIDA -- Astronomical Image Decomposition and Analysis \citep{uslenghi05}.  
The most critical part of the analysis
is the determination of the PSF model and the choice of the background level 
(that affects the faintest external signal from the object). 
In the observed field (see Figure 1) of this QSO we were able to 
use many stars to characterize the PSF model.
The fit of the target was obtained, assuming a combination
of a point source and an elliptical galaxy convolved with the proper
PSF. The radial brightness profile of the quasar compared with that 
of the PSF is shown in Figure 1. 

The result of the best fit  indicates that the host galaxy is a 
luminous  elliptical: K=19.1 $\pm$ 0.3  corresponding to M$_K$ = -27.1
and an effective radius Re $\sim$ 7 kpc. 


\begin{figure}[htbp]
    \centering
   {\includegraphics[scale=.4]{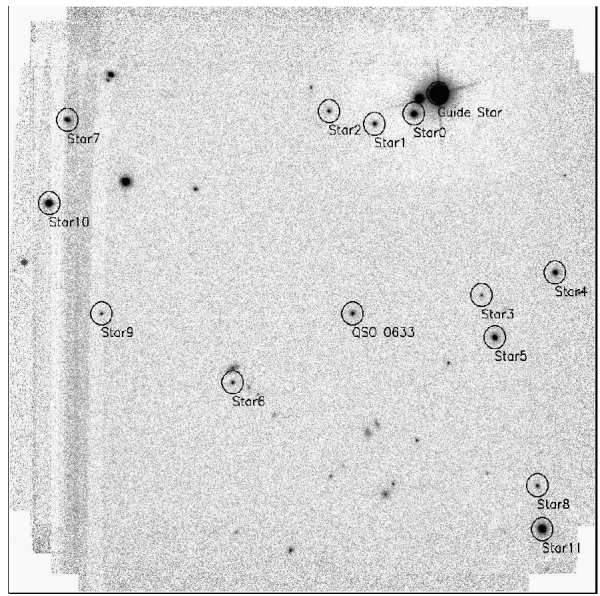}
   \includegraphics[scale=.7]{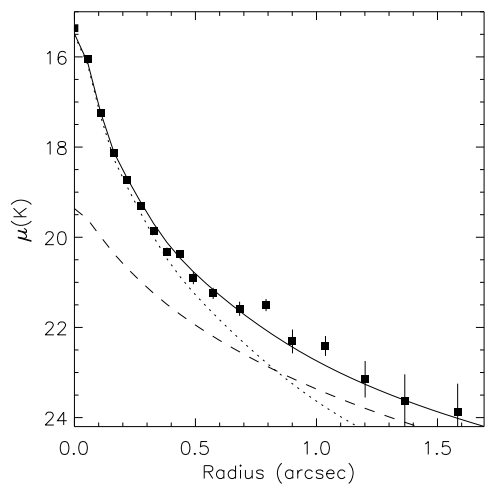} 
    }
    \caption[prof0633]{ {\it Left:} The K-band image of the radio loud quasar 
  (z = 2.928) WGA J0633.1-2333.  {\it Right:} The radial brightness profiles 
(filled squares), superimposed to the fitted model (solid line) consisting of the PSF 
(dotted line) and an elliptical  galaxy convolved with 
its PSF (dashed line). The image decomposition was performed by AIDA (see text). 
The associated errors are a combination of the statistical 
photometry in each bin and of the uncertainty on the background level. 
}
  \label{fig1}
\end{figure}

This new measurement, together 
with the available data on quasar hosts at lower redshift, are reported in Figure 2, which shows
that up to z $\sim$ 2.5, RLQ appear hosted by 
massive fully formed galaxies that are undergoing passive evolution.
The indication is that at least up to z=2.5  there is no 
 decrease in mass (luminosity) as would be expected in the hierarchical merging 
scenario for the formation and evolution of massive spheroidal galaxies 
\citep{kauffmann00}.

On the other hand the host galaxy of the RLQ  at z $\sim$ 3  is 
fainter by about a factor of 2 with respect to the (average) value that is expected if the 
RLQ hosts luminosity trend depicted at lower z continues also at higher redshift.
This may be suggestive of a real drop in luminosity and likely also in mass 
of the hosts of the quasars. 
This downturn in luminosity, that appears to occur very close to the 
epoch where there is the peak of QSO activity, if confirmed, thus 
indicates a very early epoch for the last major merger of galaxies  and emphasizes 
a strict QSO-host co-evolution.

Because of the potential cosmological importance of the result in the context 
of the models of galaxy and BH formation it is  mandatory to 
resolve a sizeable number of objects at z $\sim$ 3 and beyond. 
We have an ongoing program at VLT to explore the quasar host properties 
of a sizeable sample of z $>$ 2 objects.


\begin{center}
\begin{figure} 
  {\includegraphics*[scale=.7]{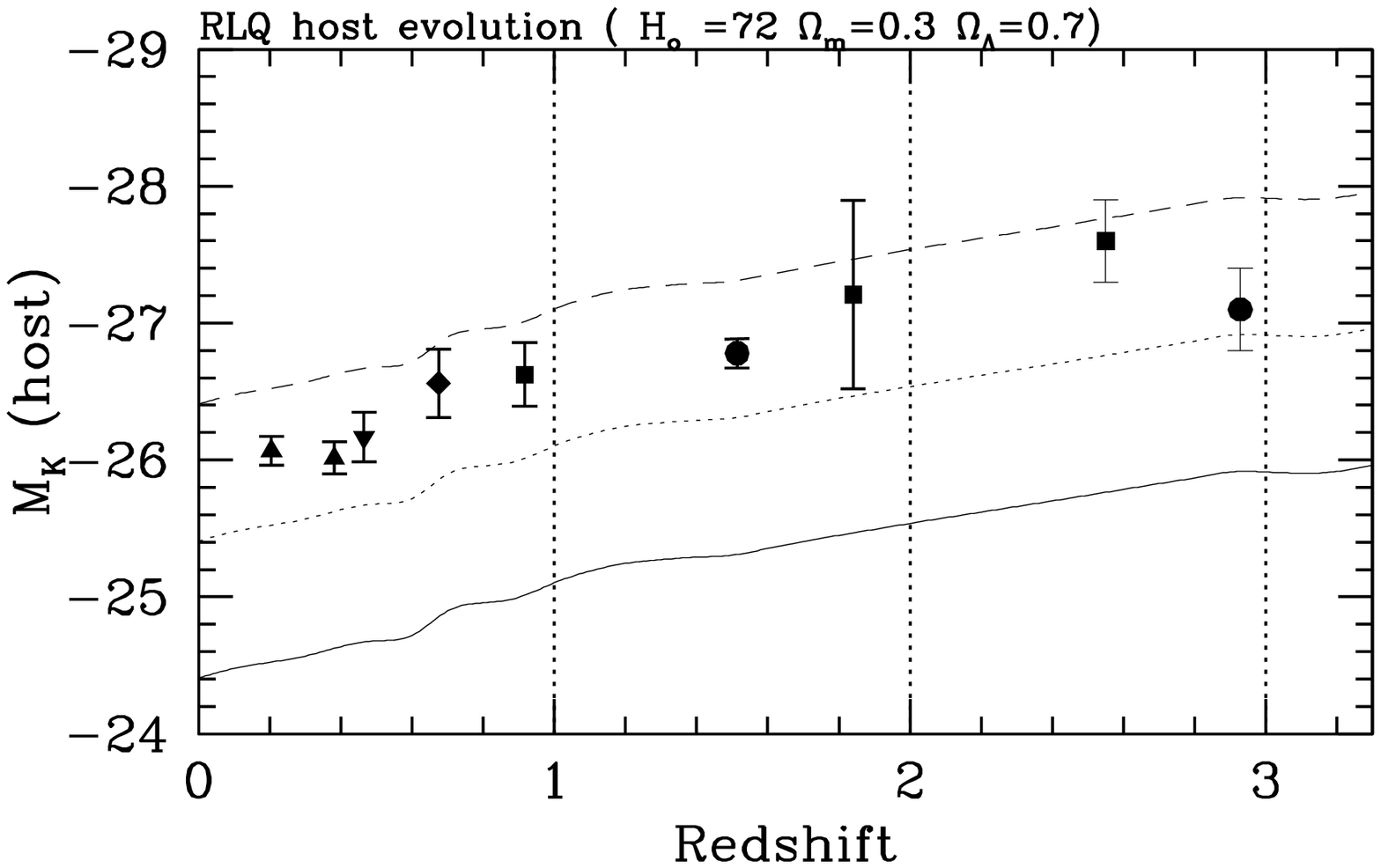} }
\caption[rlqevol]{ The evolution of radio loud quasar host luminosity
compared with that expected for massive ellipticals (at M$^*$, M$^*$-1
and M$^*$-2;  solid, dotted and dashed line ) undergoing passive
stellar evolution \citep{bressan98}.  The host galaxy of two RLQ 
at z $\sim$ 2.5 and 2.9 is compared with the data for samples of
lower redshift RLQ presented in \citet{falomo04}.  Each point is
plotted at the mean redshift of the considered sample while the error
bar represents the $1\sigma$ dispersion of the mean except for the individual objects 
at z $>$ 2 where the  uncertainty of the measurement is given.
\label{rlqevol}}
\end{figure}
\end{center}




\end{document}